\documentclass[letterpaper, 10 pt, conference]{ieeeconf}  % Comment this line out if you need a4paper

\IEEEoverridecommandlockouts                              % This command is only needed if 
\usepackage{graphics} % for pdf, bitmapped graphics files
\usepackage{epsfig} % for postscript graphics files
\usepackage{mathptmx} % assumes new font selection scheme installed
 \usepackage{amsmath,bm}
 
 \usepackage{amsfonts}
\usepackage{amssymb}
\usepackage{times} % assumes new font selection scheme installed
 \usepackage{graphicx}
\usepackage{xcolor}
\title{\LARGE \bf
 Model-Free Control Design for Feedback Linearizable
SISO Systems
}

\author{Karthik Shenoy$^{1}$, Akshit Saradagi$^{2}$, Ramkrishna Pasumarthy$^{1}$, Vijaysekhar Chellaboina$^{3}$% <-this % stops a space
\thanks{$^{1}$Karthik  and Ramkrishna  are with the Dept. of Electrical Engineering, IIT-Madras, India.
        {\tt\small ee21d405@smail.iitm.ac.in, ramkrishna@ee.iitm.ac.in}}%% <-this % stops a space
\thanks{$^{2}$Akshit is a Postdoctoral researcher at the Department of Computer Science, Electrical and Space Engineering, Lule\aa~University of Technology, Sweden. 
        {\tt\small akssar@ltu.se}}%%
\thanks{$^{3}$ Vijaysekhar  is the Dean of Engineering, GITAM (Deemed to be University), Visakhapatnam, India. 
        {\tt\small dean\_engineering@gitam.edu}}%
}

\newtheorem{thm}{Theorem}
\newtheorem{asm}{A}

\usepackage[colorlinks=false]{hyperref}
\begin{document}

\maketitle
\thispagestyle{empty}
\pagestyle{empty}

%%%%%%%%%%%%%%%%%%%%%%%%%%%%%%%%%%%%%%%%%%%%%%%%%%%%%%%%%%%%%%%%%%%%%%%%%%%%%%%%
\begin{abstract}                % Abstract of not more than 250 words. 

Data-driven control has gained significant attention in recent years, particularly regarding feedback linearization of nonlinear systems. However, existing approaches face limitations when it comes to implementing them on hardware. The main challenges include the need for very small sampling times, which strain hardware capabilities, and the requirement of an initial open-loop data set, which can be impractical for stabilizing unstable equilibrium points. To address these issues, we propose a two-stage model-free approach that combines a high-gain observer and a dynamic controller. This eliminates the hardware implementation difficulties mentioned earlier. The high-gain observer acts as a robust state estimator, offering superior noise attenuation and lower computational costs, crucial factors for digital hardware implementation. Unlike data-driven methods, our design's stability and performance depend on a tunable software parameter, simplifying digital implementation without overburdening hardware resources. Experimental results on a Twin Rotor system demonstrate the effectiveness of our approach compared to the state-of-the-art data-driven method.

%In a recent work by Fraile et al., a data-driven estimator coupled with a dynamic controller was proposed for stabilizing SISO systems that are feedback linearizable. In this article, we propose a design methodology for the model-free control of feedback-linearizable systems, using a high-gain observer and a dynamic controller.   We validate our results by implementing both techniques on a twin-rotor system and conclude that the method we present exhibits better performance in terms of system transients and noise rejection. {\em We then show how the number of samples used in estimation has an effect on the performance of a data-driven control system.} \red{What is concluded?} 
\end{abstract}

%%%%%%%%%%%%%%%%%%%%%%%%%%%%%%%%%%%%%%%%%%%%%%%%%%%%%%%%%%%%%%%%%%%%%%%%%%%%%%%%
\section{INTRODUCTION}
Traditional control techniques often rely on accurate system models 
%or designing control laws when bounds on model uncertainties are known. 
which are either based on first principles or  system identification methods, where models are derived from input-output data. With systems becoming increasingly large-scale and processes getting more complex, designing controllers using physical models (even though available) may become intractable. When rich sensor data from such systems is available, the control laws can be designed directly using the data and in the process obviating the need for physical models and explicit model identification. Model-free methods have previously been employed in control, in the form of adaptive control \cite{adaptive}, iterative learning control \cite{itfbtun}, PID tuning \cite{pidtune}, \cite{intelpid} to name a few.
 Data-driven control is yet another technique that belongs to the spectrum of model-free approaches and in the recent years, efforts have been made towards deriving traditional control laws such as state feedback \cite{formula}, MPC \cite{mpc},  LQR \cite{LQR}, minimum-energy control \cite{minen}, event-triggered control \cite{etc}, and even verifying certain dissipativity properties \cite{dissipativity} directly from the measured data, without the need of explicit model identification. The aforementioned approaches require persistently exciting data in order to uniquely identify the system and perform control design. The authors in \cite{informativity} provide a framework to deal with data that need not be persistently exciting.
 
While much of this literature has defected towards data-driven control of linear systems, very few results have been reported for the case of   nonlinear systems  \cite{nldatadriven}, \cite{NLdatadriven2}. An interesting result is presented in \cite{tabuada} for the class of feedback linearizable systems, where the authors use a data-driven estimator together with a dynamic control law.
%It is quite intriguing to see the recent works of Lucas Fraile et al. \cite{tabuada}, in which a model-based technique like feedback linearization, can be implemented without the knowledge of the plant model.\footnote{Even though the term used is model-free, it requires some knowledge on the class of systems and the design of the controller is much easier when some bounds on the plant dynamics are known.} using the measurement data collected online. 
% The results presented use a data-driven estimator along with a simple dynamic controller, for  systems that are feedback linearizable. %The results in the paper can also be extended to  partially feedback linearizable systems. 
 The dynamic controller is the key element in the design that makes the controller model-free. A limitation of the method proposed in \cite{tabuada} is the dependency of the boundedness and convergence properties of the states on the sampling time, which could place stringent requirements on the hardware used in the digital implementation of the controller. 
 In this article, we propose a model-free approach that employs a high-gain observer as the state estimator, which has no such critical dependence on a hardware parameter and offers ease of practical implementation. The main contributions of this article are summarized as follows:
 %We show how the sampling time, the number of samples used in data-driven estimation, etc play an important role in the performance of the system.
%We gained some fascinating insights on how a data-driven estimator like the one proposed in \cite{tabuada} could be replaced by a high-gain observer. Since the dynamic controller proposed in \cite{tabuada} is quite elegant, we use the same controller in conjunction with an high-gain observer which does not require any apriori measurement data, and whose estimation 
 %error convergence and boundedness depend on a tunable-parameter $\epsilon$, for the model-free control of feedback-linearizable systems.  
\begin{itemize}
\item We present a model-free approach for designing stabilizing controllers for (partially-) feedback linearizable systems, using high-gain observers. We show that the proposed method is more suitable for digital implementation and with respect to noise attenuation, compared to the recent data-driven methods \cite{tabuada}. The proposed model-free approach does not require any prior open-loop data collection, which provides it with a distinct advantage over data-driven approaches, when the hardware plants are unstable.
\item Our design approach has two key stages. The first stage is the estimation stage, which comprises of a high-gain observer, which is a robust non-linear state estimator (see \cite{khalilhgo},\cite{exthgo},\cite{lowpower}) capable of providing state estimates without taxing the hardware resources (sampling rate and computational cost). The second stage uses the dynamic controller presented in \cite{tabuada}, which is simple and elegant, as the required feedback linearizing input can be computed dynamically by adjusting just one tunable parameter (the controller gain).
\item We present experimental validation of the proposed method using a twin-rotor system, without any prior knowledge of its dynamics. The high-gain observer combined with the dynamic controller stabilizes the yaw-angle of a twin rotor, and tracks the step input robustly, in the presence of large sensor noise.
\item We compare the method proposed in the article with the data-driven technique in \cite{tabuada} experimentally, in terms of the hardware resources utilized (sampling rate and computational costs) and robustness in the presence of sensor noise, which are crucial in hardware digital implementation. The computational cost for the high-gain observer stage is shown to be far lower compared to the data-driven estimator in \cite{tabuada}, while achieving the same performance and noise attenuation.
 
%\item 
   % \item The stabilization of a non-linear feedback-linearizable system, whose model is unknown, without the use of any data has been done. We should focus on the discrete-time implementation as well.
    %\item We critically review results related to two model-free approaches and present our findings by assessing 1) The claim as a model-free approach : Although claimed, some stages of the design assume knowledge about the model. 2) Ease of translation to digital implementation 3) Performance in the presence of noise 4) Approximations made and the ease of analysis.
   % \item We present our perspective on the popular approaches for model-free control specifically focusing on the feedback linearization problem.
  %  \item Our critical review of the merits and the shortcoming of both the methods leads us to a formulation of a problem whose solution that is presented in this paper. The solution incorporates the simplicity and elegance of the discrete-time dynamic controller from Tabuada's work and the ability of High-gain observers to provide good estimates of the states without taxing and stretching the hardware resources.
   % \item The gap that existed between theory and practice when using the data-driven estimator has been bridged. 
\end{itemize}
The rest of the article is organized as follows: In Section \ref{pab} we briefly present the concepts of feedback linearization and the high-gain observer. In Section \ref{SMCD}, the approximate discrete-time plant models are derived and a  stabilizing control law is designed. We combine the dynamic controller and the high-gain observer and present the main results of the article in Section \ref{MR}. In Section \ref{discuss} we critically review our results, comparing it with the results proposed in \cite{tabuada}. % We then compare both techniques and show the flexibility of using a high-gain observer using simulations in Section \ref{sim}. 
 Experimental validation and comparison are provided in Section \ref{exp} using a twin-rotor MIMO system.
\section{PRELIMINARIES AND BACKGROUND}
\label{pab}
\subsection{Notation}
$\mathbb{R}$ represents the set of real numbers and $\mathbb{R}^+,\mathbb{R}^{++}$ represents the set of non-negative and positive real numbers respectively. $\mathbb{R}^n$ is the space of all real $n$-dimensional vectors. All the estimated states are represented with a hat on top, for example, the estimate of state $x$ as $\hat{x}$. $\mathrm{col}\{x_1,\dots, x_n\}$ would represent the $n\times1$ column vector with entries $x_1\dots x_n$ and $\mathrm{diag}(x_1,\dots,x_n)$ represents a diagonal matrix with the entries $x_1,\dots x_n$. The 2-norm of a vector will be represented by $\left\|x\right\|$. The Lie-derivative of a function $h(x): \mathbb{R}^n\rightarrow\mathbb{R}$ along a vector field $f(x):\mathbb{R}^n\rightarrow\mathbb{R}^n$ is given by $L_fh(x)=\frac{\partial h(x)}{\partial x}\cdot f(x)$. The big-$O$ notation given by $f(x)=O_x(K)$ implies there exists some $H,K\in\mathbb{R}^+$ such that for all $k\in[0,T]$, we have $\left\|f(x,k)\right\|\leq HK\left\|x\right\|$. $\lambda_{\mathrm{min}}(A)$ represents the smallest eigenvalue of $A$.

\subsection{Feedback Linearization}
Feedback linearization is a technique in which a nonlinear system can be transformed into a linear system via a proper choice of a nonlinear transformation and a  state feedback control law.
%Feedback linearization is a method in nonlinear control which revolves around the idea of transforming the nonlinear system dynamics into a linear one, fully or partly, by the means of state transformations and feedback. 
Feedback linearization techniques have found wide variety of applications, especially  in the control of aerospace and robotic systems.  More literature on feedback linearization can be found in \cite{slotine} and \cite{isidori}. Consider the single-input-single-output nonlinear system in its normal form:
\begin{equation}\label{trfflsys1}
    \begin{aligned}
    \dot{w}&=f_0(w,x) \\
        \dot{x}&=A_cx+B_c[a(w,x)+b(w,x)u]\\ 
    y&=C_cx        
    \end{aligned}
\end{equation}
where $x\in\mathbb{R}^{\rho}$, $\rho$ is the relative degree of the system for a given output. $w$ are the internal states, which are not observable from the given output. $a(x,w)$ and $b(x,w)$ are nonlinear functions from $\mathbb{R}^n\rightarrow\mathbb{R}$ (which later on are assumed to be unknown). $A_c,\;B_c,\;C_c$ are in the Brunovsky canonical form representation of a chain of $\rho$ integrators. Now by choosing a control law of the form $u={b(w,x)}^{\mbox{--}1}(v(x)-a(w,x))$, where $v(x)$ is any linear control law, we can linearize the input-output dynamics to obtain:
\begin{equation}\label{fblinsys}
    \begin{aligned}
         \dot{w}&=f_0(w,x)\\
    \dot{x}&=A_cx+B_cv(x)\\
    y&=C_cx.
    \end{aligned}
\end{equation}
The zero dynamics of \eqref{fblinsys} is given by $\dot{w}=f_0(w,0)$.
We assume that the zero dynamics of the system is asymptotically stable. This is to ensure that the internal dynamics, which is uncontrollable, is asymptotically stable.
\subsection{The High-Gain Observer}\label{HGOsect}
The high-gain observer, presented in  \cite{khalilhgo}, \cite{exthgo}, \cite{lowpower}, is a non-linear robust state estimator. The main features of the high-gain observer are that the estimation errors decay rapidly towards small values. Additionally, it is quite robust with respect to model uncertainties. The dynamics of the high-gain observer, for system \eqref{trfflsys1}, is defined as follows:
\begin{align*}
\dot{\hat{x}}=A\hat{x}+B\phi_0(\hat{x},u)+H(y-C\hat{x})
\end{align*}
where $\hat{x}\in\mathbb{R}^{\rho}$ and $H=\mathrm{col}\{
   \alpha_1/\epsilon,\;
    \alpha_2/\epsilon^2,\;
    \hdots,
    \alpha_{\rho}/\epsilon^{\rho}\}$, $\phi_0(\hat{x},u)$ is a nominal model for $a(w,x)+b(w,x)u$ and is locally Lipschitz in $(x,u)$ over the domain of interest and is globally bounded in $x$. $\epsilon\in\mathbb{R}_0^+$ and positive constants $\alpha_i$ are chosen such that the polynomial:
 \begin{align}\label{polynomial}
     s^{\rho}+\alpha_1s^{\rho-1}+\dots+\alpha_{\rho-1}s+\alpha_{\rho}
 \end{align}
 is Hurwitz. Note that $\epsilon$ is the observer time constant.
 
 The high-gain observer can also be used as a disturbance estimator, by treating the disturbance as an additional state. The dynamics of the extended high-gain observer is given by:
 \begin{align*}
     \begin{bmatrix}\dot{\hat{x}}\\\dot{\hat{x}}_{\rho+1}\end{bmatrix}&=\bar{A}\begin{bmatrix}\hat{x}\\\hat{x}_{\rho+1}\end{bmatrix}+\bar{B}\phi_0(\hat{x},\hat{x}_{\rho+1},u)\nonumber\\&+\bar{H}(y-\bar{C}\begin{bmatrix}\hat{x}\\\hat{x}_{\rho+1}\end{bmatrix})
 \end{align*}
 where $\hat{x}_{\rho+1}$ is the additional state representing the uncertain disturbance input and $\bar{A},\bar{B},\bar{C}$ the canonical representation of a chain of $\rho+1$ integrators. Here, the gain matrix $\bar{H}$ is : 
    \begin{align*}
     \bar{H}=\begin{bmatrix}
   H\\
\alpha_{\rho+1}/\epsilon^{\rho+1}\\
    \end{bmatrix}.
\end{align*}
In order to discretize the high-gain observer, we first perform a change of coordinates:
\begin{align*}
q=D\begin{bmatrix}\hat{x}\\\hat{x}_{\rho+1}\end{bmatrix}=D\hat{\bar{x}}
\end{align*}
where the matrix $D$ is a transformation matrix given by $D=diag(1,\epsilon,\epsilon^2,\dots,\epsilon^{\rho})$. This transformation is made so as to remove the negative powers of $\epsilon$ from the gain matrix $\bar{H}$ before discretizing the observer. 

Upon discretizing the observer dynamics using the bi-linear transformation methods  with sampling time $T$ and choosing $\phi_0=0$ we obtain:
\begin{equation}\label{obsv}
\begin{aligned}
    \xi(k+1)&=A_d\xi(k)+B_dy(k)\\
    \hat{\bar{x}}(k)&=D^{\mbox{--}1}[C_d\xi(k)+D_dy(k)]
\end{aligned}
\end{equation}
where the observer state $\xi\in\mathbb{R}^{\rho+1}$ and the matrices $A_d,B_d,C_d$, and $D_d$ are the coefficients of the discrete-time implementation of the high-gain observer, given in Table 9.1, \cite{khalilhgo}. The relation between the sampling time $T$ and the observer time constant $\epsilon$ is:
\begin{align}
    T=\alpha\epsilon,\;\alpha\in\mathbb{R}_{>0}. \label{Tepsalph}
\end{align}
Any value of $\alpha>0$ could be chosen in the bi-linear transformation case, as the matrix $A_d$ is Hurwitz for all positive values of $\alpha$. Theorem 9.1 in \cite{khalilhgo} guarantees the existence of an $\epsilon$ such that the output-feedback control law, with the state estimates obtained using the high-gain observer, stabilizes the system. The theorem is as follows:
%A small caveat here is that, $\alpha$ relates the hardware parameter $T$ to the observer time constant $\epsilon$ and in general, $T<\epsilon$. Hence we may not be able to reduce $T$ to really small values since there would be hardware limitations. Hence in such cases $\alpha$ would be kept close to $1$. 
\begin{thm}\label{hgothm}
(Theorem 9.1, \cite{khalilhgo})
Consider the closed-loop system with the plant \eqref{trfflsys1} and the output
feedback discrete controller $u(\hat{\bar{x}}(k))$ with the observer \eqref{obsv}. Let $\mathcal{R}$ be the region of attraction of the system \eqref{trfflsys1} with the controller $u(\bar{x}(k))$, and $\mathcal{S}$ be any compact set in the interior of $\mathcal{R}$, and let $\mathcal{D}$ be any compact subset of $\mathbb{R}^{\rho+1}$.  Suppose $((w(0),\bar{x}(0)), \hat{\bar{x}}(0))\in\mathcal{S}\times\mathcal{D}$. Then
\begin{itemize}
    \item there exists $\epsilon_1^*>0$ such that for every $\epsilon\in(0,\epsilon_1^*]$, $(w(t),x(t))$ is bounded for all $t\geq0$ and the estimation error $e_{\Tilde{\bar{x}}}(k)$ is bounded for all $k\geq0$
    \item given any $\mu>0$, there exists an $\epsilon_2^*>0$, $T_1>0$ and $k^*>0$ all dependent on $\mu$ such that for every $\epsilon\in(0,\epsilon_2^*]$,
    \begin{align*}
        &\left\|(w(t),x(t))\right\|\leq\mu\;\forall t\geq T_1\\
        &\left\|e_{\Tilde{\bar{x}}}(k)\right|\leq\mu\;\forall k\geq k^*
    \end{align*}
    
    %\item given anny $\mu>0$, there exists an $\epsilon_3^*>0$ dependent on $\mu$, such that for every $\epsilon\in(0,\epsilon_3^*]$,
    %\begin{align*}
     %   \left\|(w(t),x(t))\right\|
    %\end{align*}
    
    \item If the origin of the system \eqref{trfflsys1} with the controller $u(\bar{x}(k))$ is exponentially stable and $f_0(w,x)$, $Ax+B\phi(w,x,u)$
        are twice continuously differentiable in the neighborhood of the origin, then there exists an $\epsilon_4^*>0$ such that for every $\epsilon\in(0,\epsilon_4^*]$, the origin of \eqref{obsv} and the discretized plant \eqref{fblinsys} is exponentially stable and $\mathcal{S}\times\mathcal{D}$ is a subset of its region of attraction. Moreover, the continuous-time trajectory $(w(t),x(t))$ decays to zero exponentially fast.
\end{itemize}

\end{thm}
\section{SYSTEM MODEL AND CONTROLLER DESIGN}  \label{SMCD}
In this section, we derive approximate discrete-time models of the plant and discuss the controller design, which incorporates the observer time constant $\epsilon$. 
\subsection{System Model} 
Before getting into the discretized models, we show $O_x(T)=O_x(\epsilon)$ when in \eqref{Tepsalph},  $\alpha\leq1$:
\begin{align}\label{OTOe}
    \left\|f(x,t)\right\|&\leq HT\left\|x\right\|=H\alpha\epsilon\left\|x\right\|\leq H\epsilon\left\|x\right\|
\end{align}
where the first equality is obtained using \eqref{Tepsalph} and the third inequality is obtained by setting $\alpha\leq1$.

For the ease of deriving the control law, we set the relative degree $\rho=2$. The same procedure would follow for systems with a higher relative degree. We also assume that the measurement is free of noise. Hence we obtain the simplified model in the below normal form:
\begin{align}
    &\dot{w}=f_0(w,x) \label{mainressys1}\\
    &\dot{x}_1=x_2\nonumber\\
&\dot{x}_2=a(w,x)+b(w,x)u\label{mainressys4}\\
    &y=x_1 \nonumber
\end{align}
where $w\in\mathbb{R}^l$ and $x\in\mathbb{R}^{2}$ constitute the state vector. We make the following assumptions:
\begin{asm}
$u(kT+\lambda)=u(kT), \forall\lambda\in[0,T)$ where $T$ is the sampling time.\label{asm1}
\end{asm}

\begin{asm}For a small enough sampling time $T$, we assume $a(w,x)$, $b(w,x)$ to be constants between two consecutive sampling instants. \label{asm3}
\end{asm}

\begin{asm}
The origin $w=0$ of the zero dynamics $\dot{w}=f_0(w,0)$ is asymptotically stable. \label{asm4}
\end{asm}
Consider the subsystem \eqref{mainressys4}. On discretizing this system using Taylor-series expansion, and using \eqref{OTOe}, we  obtain the exact discretization $F^e(x,u)$:
\begin{equation}
    \begin{aligned}
x_1(k+1)&=x_1(k)+\epsilon\alpha x_2(k)+\frac{{(\epsilon\alpha)}^2}{2}(a(k)+b(k)u(k))\nonumber\\&+O_{(x,u)}(\epsilon^3) \\
x_2(k+1)&=x_2(k)+\epsilon\alpha(a(k)+b(k)u(k))+O_{(x,u)}(\epsilon^2)\\
     y(k)&=x_1(k).
\end{aligned}       
\end{equation}
 Neglecting the $O(\epsilon^2)$ terms, we obtain the approximate dynamics $F^a(x,u)$. Since $a(w,x)$ and $b(w,x)$ are unknown, we take $a(w,x)+b(w,x)u(k)$ as an extended state $x_{3}(k)$. Hence the approximate discretized extended system dynamics are:
\begin{equation*}
\begin{aligned}
\bar{x}(k+1)&=\begin{bmatrix}
 1&\epsilon\alpha&\frac{{(\epsilon\alpha)}^2}{2}\\
 0&1&\epsilon\alpha\\
 0&0&1
\end{bmatrix}\bar{x}(k),\;\;y(k)=C\bar{x}(k)
    \end{aligned}    
\end{equation*}
where $C=\begin{bmatrix}
 1&0&0
\end{bmatrix}$.
\subsection{Controller Design}\label{contsect}
\subsubsection{Linear-Part of the Controller}
In this section we present the design of the stabilizing controller for the feedback linearized system, assuming we have the knowledge of $a(k)$ and $b(k)$. A feedback linearizing control law of the form:
\begin{align}\label{fblincontrol}
    u(k)&=b^{-1}(k)(-a(k)+v(x(k))).
\end{align}
Substituting \eqref{fblincontrol} in the approximate model $F^a(x,u)$, we obtain $   x(k+1)=Ax(k)+Bv(k)$, where 
\begin{equation*}
\begin{aligned}
    A&=\begin{bmatrix}
     1&\epsilon\alpha\\0&1
    \end{bmatrix} =I+A_i\epsilon\alpha\\ 
    B&=\begin{bmatrix}
     \dfrac{{(\epsilon\alpha)}^2}{2}\\\epsilon\alpha
    \end{bmatrix}=B_i\epsilon\alpha+B_j{(\epsilon\alpha)}^2.
\end{aligned}    
\end{equation*}

Since $(A_i,B_i)$ is a controllable pair, we can find a control law $v(x)=Kx$ and symmetric matrices $P_x\succeq0$ and $Q\succ0$ such that:
\begin{align*}
{(A_i+B_iK)}^TP_x+P_x(A_i+B_iK)=-Q
\end{align*}
and $V_x(x)=x^TP_xx$. For the approximate dynamics, we can show that:
\begin{align}\label{ineq1}
    V_x(F^a(x,u))-V_x(x)\leq-\lambda_{min}(Q)\epsilon{\left\|x\right\|}^2+O_{{(x,u)}^2}(\epsilon^2).
\end{align}
\subsubsection{Dynamic Controller}
The stabilizing controller is designed assuming we have the knowledge of $a(k)$ and $b(k)$. Since we don't have that knowledge, we need to design a dynamic controller such that the error, $e_u$ defined by:
\begin{align}
    e_u=v(x(k))-x_3(k)
\end{align}
asymptotically converges to the origin. We use the dynamic control law proposed in \cite{tabuada} for this purpose. The dynamic control with state feedback is:
\begin{align}\label{dyncontrollawsttefb}
    u(k+1)=u(k)+\gamma(v(x(k))-x_3(k))
\end{align}
where the value of $\gamma$ must be chosen such that $\gamma<\bar{b}^{\mbox{--}1}$ and $\bar{b}$ is the upper bound of $b(w(k),x(k))$. 
\section{Main Results}\label{MR}
We are now ready to state the main result in this article. Given a system of the form \eqref{trfflsys1},  which can be transformed into a (partially-) feedback-linearizable normal form, with asymptotically stable zero dynamics, the objective is to design a feedback control law, which is easy to implement, that can asymptotically stabilize the system. We also aim to eliminate the need of a finite set of measurements for state estimation. To meet this objective we make use of high-gain observers, which have been extensively used to design output feedback controllers for nonlinear systems. We will show that our results guarantee the existence of a high-gain observer time constant $\epsilon$, such that the combination of the high-gain observer with the dynamic controller, along with feedback linearization drives the system states to the origin asymptotically and keeps all the signals bounded under zero measurement noise conditions. 

\begin{thm}\label{th1}
 Consider the nonlinear system of the form \eqref{mainressys1}, together with the assumptions \hyperref[asm1]{A1}-\hyperref[asm4]{A3}, where the functions $f_0(w,x),a(w,x)$ and $b(w,x)$ are unknown. The initial conditions $((w(0),x(0)),\hat{x}(0))\in\mathcal{S}\times\mathcal{D}$, where the sets $\mathcal{S}\subset\mathbb{R}^{l+\rho}$ and $\mathcal{D}\subset\mathbb{R}^{\rho}$ are compact. Then, in the absence of measurement noise, the extended high-gain observer \eqref{obsv}, and the dynamic controller \eqref{dyncontrollawsttefb}, guarantees the existence of an $\epsilon^*\in\mathbb{R}^+$, and constants $b_1>0$, $T_1>0$, $k^*>0$  such that for any $\epsilon\in(0,\epsilon^*]$, the closed loop trajectories are bounded, i.e.
\begin{align*}
    &\left\| \hat{x}(k)\right\|\leq b_1,\;\forall k\geq k^*\\
    &\left\| x(t)\right\|\leq b_1,\;\forall t\geq T_1\\
    &\left\| w(t)\right\|\leq b_1,\;\forall t\geq T_1\\
    &\left\| e_u\right\|\leq b_1.
\end{align*}

{\it Moreover, $\displaystyle{\lim_{t \to \infty}x(t)=0}$ and $\displaystyle{\lim_{t \to \infty}w(t)=0}$}.
\end{thm}
\textbf{Proof}
The proof will consist of two parts. In the first part of the proof need to show how a state-feedback control law guarantees the stability of the closed-loop system. This will be shown using similar techniques used in the proof of Theorem 8.1 in \cite{tabuada}, in addition to using the relation derived in \eqref{OTOe}. Next, we need to prove that the output-feedback controller stabilizes the system, with estimates obtained using the high-gain observer. 

Consider the Lyapunov candidate function $\bar{V}=V_x(x(k))+V_e$, where $V_x=x^TP_xx$ as given in section \ref{SMCD}, and $V_e=\dfrac{1}{2}e_u^2$ for the error dynamics. Let $\mathcal{R}$ be the smallest sub-level set of $\bar{V}$.  We show that $\mathcal{R}$ is an invariant set using the steps (1)-(4) below. 
\begin{enumerate}
    \item First we relate the evolution of function $V_x$ along the trajectories of the exact discretized system $F^e(x,u)$ with the evolution of the approximate dynamics $F^a(x,u)$ upto $O(\epsilon^2)$ terms. Since $u=b^{\mbox{--}1}(k)(\mbox{--}a(k)+v(x))\mbox{--}b^{\mbox{--}1}(k)e_u(k),\;\left\|(x,u)\right\|\leq\left\| x\right\|+\left\| b^{\mbox{--}1}(k)(-a(k)+v(x))\right\|+\left\| b^{{\mbox{--}1}}(k)e_u(k)\right\|,\;$
     and \\$\left\| b^{\mbox{--}1}(k)(-a(k)+v(x))\right\|$ is Lipschitz continuous in $x$, we arrive at the following inequality:
    \begin{align}\label{approxexactlyap}
         V_x(F^e(x,u))-V_x(x)&\leq V_x(F^a(x,u))-V_x(x)\nonumber\\&+O_{{x}^2}(\epsilon^2)+O_{{e_u}^2}(\epsilon^2).
    \end{align}
    \item Next we show that $V_x(F^a(x,u))-V_x(x)$ is a negative definite quantity upto $O(\epsilon^2)$ terms using \eqref{ineq1} and choosing $\lambda_x\leq\frac{\lambda_{min}(Q)}{2}$ and by completing the squares:
    \begin{align}\label{lyap1}
        V_x(F^a(x,u))-V_x(x)&\leq-\lambda_x\epsilon {\left\| x\right\|}^2+O_{{x}^2}(\epsilon^2)\nonumber\\&+O_{{e_u}^2}(\epsilon^2).
    \end{align}
    
    \item The controller error dynamics, approximated upto $O_{(x,u)}(\epsilon)$ terms is given by:
    \begin{align}\label{errordyn}
        e_u(k+1)=(1-b(k)\gamma)e_u(k)+O_{(x,u)}(\epsilon).
    \end{align}
    \item Now we show $V_e(k+1)-V_e(k)$ is negative definite up to $O(\epsilon^2)$ terms, by completing the squares, and by choosing a small enough $\gamma,\lambda_u\in\mathbb{R}^+$:
    \begin{align}\label{lyap2}
        V_e(k+1)-V_e(k)&\leq\mbox{--}\lambda_ue_u^2+O_{{x}^2}(\epsilon^2)\nonumber\\&+O_{{e_u}^2}(\epsilon^2).
    \end{align}
    Combining \eqref{approxexactlyap}-\eqref{lyap2}, we have the result:
    \begin{align*}
        \bar{V}(F^e)-\bar{V}(k)&\leq-\lambda_x\epsilon {\left\| x\right\|}^2\mbox{--}\lambda_ue_u^2+O_{{x}^2}(\epsilon^2)\nonumber\\&+O_{{e_u}^2}(\epsilon^2)\\
        &=-\lambda_x\epsilon {\left\| x\right\|}^2\mbox{--}\lambda_u{\left\| e_u\right\|}^2\nonumber\\&+H\epsilon^2{\left\| x\right\|}^2+H\epsilon^2{\left\| e_u\right\|}^2.
    \end{align*}
    Choosing
    \begin{align*}
     \lambda&<\min\{\lambda_x,\lambda_u\},\;
     \epsilon^*<\min\{\dfrac{\lambda_x-\lambda}{H},\dfrac{\lambda_x-\lambda}{H}\}
    \end{align*}
    we have:
    \begin{align*}
         \bar{V}(F^e)-\bar{V}(k)&\leq-\lambda\epsilon^* {\left\| x\right\|}^2\mbox{--}\lambda {\left\| e_u\right\|}^2.
    \end{align*}
    \item Thus $\mathcal{R}$ remains invariant, and the compactness of $\mathcal{R}$ implies the trajectories are bounded, i.e. $\exists\; b_2>0$ such that $\left\| x(k)\right\|\leq b_2$ and $\left\| e_u(k)\right\|\leq b_2\;\forall k\in\mathbb{N}$. Moreover the converge to the origin asymptotically.  Now, since $\mathcal{R}$ is invariant, and the dynamics are smooth, using Theorem 1 in \cite{sampleddatasys}, we can conclude that $\exists\;b_3>0$ such that $\left\| x(t)\right\|\leq b_3$. Furthermore, $\lim_{t\rightarrow\infty}x(t)=0$. So we choose $b_1=\max\{b_2,b_3\}$. This proves that the result holds for a state-feedback dynamic control law \eqref{dyncontrollawsttefb}.
\item From the assumption that the zero-dynamics of the system is asymptotically stable, the system \eqref{mainressys1}-\eqref{mainressys4} is asymptotically stable with state-feedback dynamic control law \eqref{dyncontrollawsttefb}. 
\item Now for the second part of the proof,  we can invoke Theorem \ref{hgothm}. As Theorem \ref{hgothm} guarantees that the states, as well as their estimates, are bounded, and the states $x(t),w(t)$ converge to the origin asymptotically. 
\end{enumerate}

\section{Discussion}\label{discuss}
The reader might be tempted to draw parallels between Theorem \ref{th1} of this article, Theorem 8.1 in \cite{tabuada} and Theorem 9.1 in \cite{khalilhgo}. It is remarked in \cite{tabuada}, that the least-squares-based state estimator proposed in \cite{tabuada} conceptually resembles a linear high-gain observer converging in finite-time, where the sampling parameter $T$ appearing in the design of the state estimator plays the role of $\epsilon$ in the high-gain observer, thus rendering both the methods sensitive to measurement noise. We now discuss in detail the similarities and differences in these approaches and in the process note the advantages and drawbacks of each of the methods, which opens up avenues for further research in this direction.
%\subsection*{Are the approaches completely model free?}
% The first thing to notice is that both methods aren't completely model-free. Deciding the value of the parameter $\gamma$ in the dynamic controller is much easier if the upper bound on $b(w,x)$ is known.
 
 \subsection*{Ease of implementation}
  From the implementation point of view, it is important to note that the sampling time $T$ is a hardware parameter and might pose challenges in the physical realization of methods proposed in \cite{tabuada}, even though an appropriate choice of sampling time $T \in [0,T^*]$ guarantees that the system trajectories asymptotically converge to the origin. This comes across as a gap between theory and practice. We highlight this using experiments on a twin-rotor system later in the article.\\ The $\epsilon$ in the high-gain observer presents itself as a tunable software parameter, thus offering more flexibility in implementation by allowing the designer to choose the observer time-constant $\epsilon$, and then scaling it using $\alpha$, to match the sampling time $T$.  Hence the results presented in Theorem \ref{th1} can be considered as an attempt in shortening the gap between theory and practical implementation.   
% Next, the parameter $T$, does not appear in Theorem \ref{th1} explicitly, thus the boundedness, and the convergence properties of the system is independent of an hardware parameter, making it much easier to implement. The designer does get some flexibility on choosing the observer time-constant $\epsilon$, and then scaling it using $\alpha$, to match the sampling time $T$. Where as, the value of $T^*$ that guarantees the asymptotic stability of the states in the data-driven method, may not be physically realizable. This widens the gap between theory and practice. Hence Theorem \ref{th1} can be considered as an attempt in shortening that gap.   
\subsection*{ The peaking phenomenon of the high-gain observer}
A characteristic of the high-gain observer is that the state estimates could peak to a negative power of $\epsilon$ before they decay to $O(\epsilon)$ values quickly. This is referred to as the peaking phenomena of the high-gain observer.  The combination of peaking phenomena and certain nonlinearities in the system can lead to a finite escape time. There are ways in which one can tackle this issue, one of them is by saturating state estimates or the control input $u(t)$ outside the compact set under consideration, (see \cite{exthgo}). This can also be dealt with using the low-power version of the high-gain observer, (see \cite{lowpower}).
\subsection*{The noisy case}
The extension of Theorem \ref{th1} to the case where measurement noise is present, is not discussed in this article, since both the high-gain observer as well as the data-driven estimator give the same performance in the presence of bounded measurement noise. The states can be shown to converge to a bounded set, whose bounds depend on the negative powers of $T$ in the data-driven state estimation case and the negative powers of $\epsilon$ in the high-gain observer. This again means that the designer has some control over the bounds in the case of a high-gain observer as compared to using a data-driven estimator. These similarities as well as differences are demonstrated in the next section using experiments on a twin-rotor MIMO system. 
\section{Experimental Work} \label{exp}
\subsection{The Setup}
\begin{figure}
      \centering
      \includegraphics[scale=0.08]{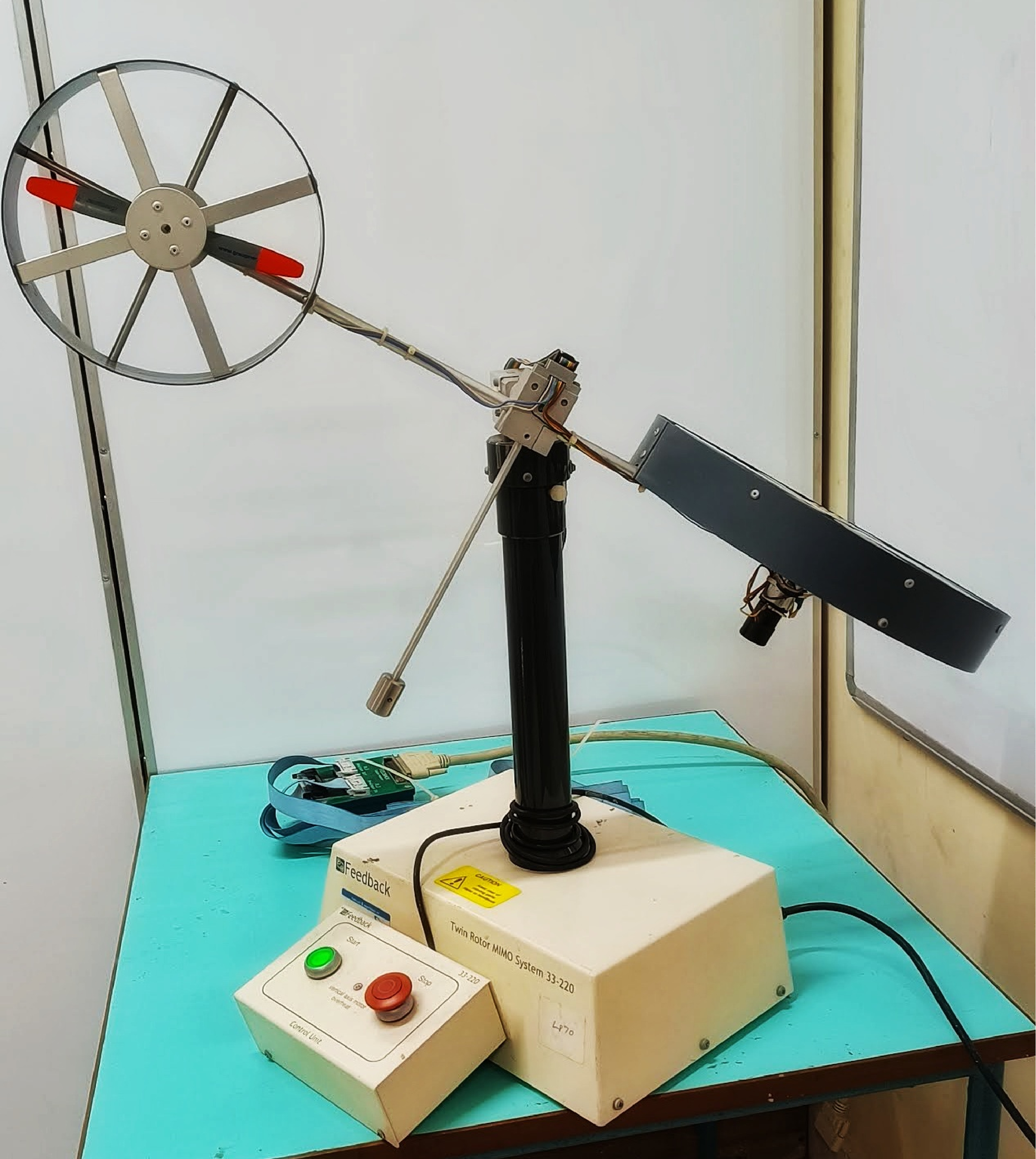}
      \caption{The experimental  Twin-Rotor test setup.}
      \label{setup}
 \end{figure}

To  analyze and compare the proposed method with the data-driven technique, we implement both methods on a twin-rotor MIMO system(TRMS). The objective is to control the yaw angle using feedback linearization by actuating the tail rotor. The pitch is left unactuated. The twin-rotor setup used is shown in Fig~\ref{setup}. The setup is run using Feedback Instruments' Simulink package. The experiments are conducted in a model-free approach, i.e., we do not have access to the mathematical model of the system. %Since our results are based on a model-free approach we do not explicitly make use of the model of the system. {\em In order to maintain the model-free nature of the results, we do not derive the governing equations of motion for the TRMS.} 
The system has a relative degree $\rho=2$. We start by designing a high-gain observer and a dynamic controller discussed in Sections  \ref{HGOsect} and \ref{contsect} respectively for the twin-rotor system.  
\subsection{High-Gain Observer and Dynamic Controller Design}\label{hgodyncde}
\begin{itemize}
    \item \textbf{Sampling Rate(T)}: All the experiments are run at a sampling rate of $T=0.001s$. 
    \item \textbf{Observer Time-Constant($\epsilon$):} As we cannot find the value of $\epsilon^*$, we choose an arbitrary initial value for $\epsilon$ and reduce it by an order of $10$, until it drops below the unknown $\epsilon^*$, that is guaranteed by Theorem \eqref{th1}. During this time, the changes in the yaw angle are made physically by giving small disturbances to the tail rotor. No control input can be given to the motor during this period since the system may turn unstable in the closed loop, when the value of $\epsilon$ is above the threshold value $\epsilon^*$(Refer Theorem \eqref{th1}). A value of $\epsilon=0.05$ was found to be good enough to estimate the states and the extended state. Thus, from \eqref{Tepsalph}, the value of $\alpha=T/\epsilon=0.02$.
    \item \textbf{Polynomial Coefficients} As the relative degree is $\rho=2$, the high-gain observer polynomial corresponding to \eqref{polynomial}, will be of degree 3 (including the extended state). Thus, in order to make the polynomial $s^3+\alpha_1s^2+\alpha_2s+\alpha_3=0$ Hurwitz, the value of $\alpha_1,\;\alpha_2,\;\alpha_3$ were chosen as $\alpha_1=6,\;\alpha_2=11,\;\alpha_3=6$.
    \item \textbf{Controller Parameter $\gamma$:} Since the model parameters are assumed to be completely unknown, the bound on the diffusion term $b(w,x)$ is not available. In order to ensure the stability of the dynamic controller error dynamics, we need $\gamma<{\bar{b}}^{-1}$ where $\bar{b}$ is an upper bound of $b(k)$. Hence to find a stabilizing $\gamma$, it was set to an initial value and reduced by orders of $10$ till the error dynamics asymptotically approached zero. A value of $\gamma=0.0003$ was found to be appropriate for the system.
    \item \textbf{Pole placement controller:} For the pole placement design, the gains were chosen as $k_1=2,\;k_2=4$. Since the system had a steady-state error, an integrator was incorporated to get the required steady-state performance. The integrator gain was set to $K_i=0.001$. To reduce the effects of peaking, the angular velocity output of the observer was saturated to a value of $\pm0.05$. Since the control input gets saturated by the inbuilt TRMS control output block, no additional saturation blocks were used to saturate the peaking in the control signal.
\end{itemize}
        
\subsection{Experimental Results}\label{results}
\begin{figure}
      \centering
    \includegraphics[scale=0.45]{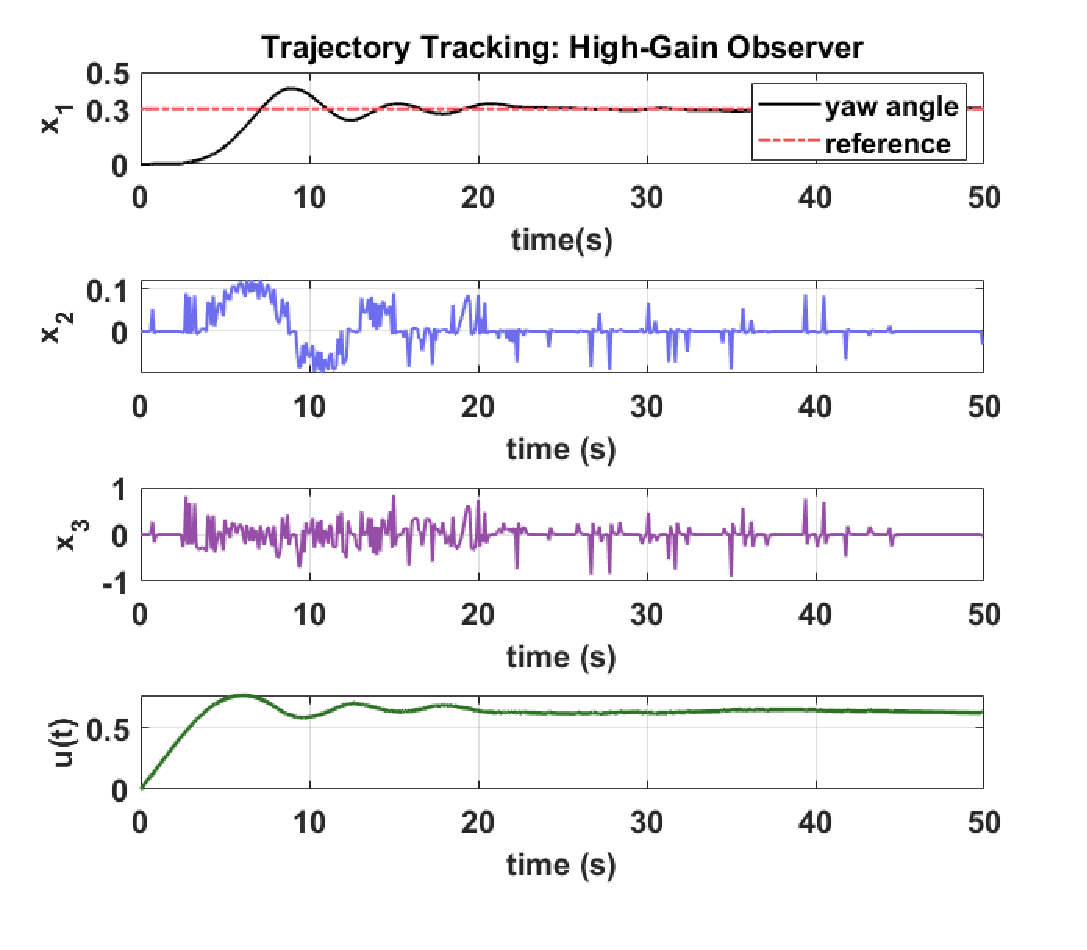}
      \caption{Plot depicting the state estimates, the extended state, and the control input, when using a high-gain observer \eqref{obsv} along with a dynamic controller.}
      \label{HGO_dyn}
 \end{figure}

The system was made to track a step input of amplitude $0.3$ rad. The selected values of $\gamma,\;K_1,\;K_2$ and $k_i$ resulted in an overshoot of $36\%$. The peaking phenomena can be seen in the angular velocity as well as the estimate of the extended state. Observer peaking does not affect the system states as the TRMS bandwidth is much lower compared to the peaking decay rate. Fig~\ref{HGO_dyn} shows the state estimate trajectories as well as the control input. 
\subsection{Comparison: Data-Driven vs High-Gain Observer}\label{compare}
\begin{figure}
      \centering
      \includegraphics[scale=0.45]{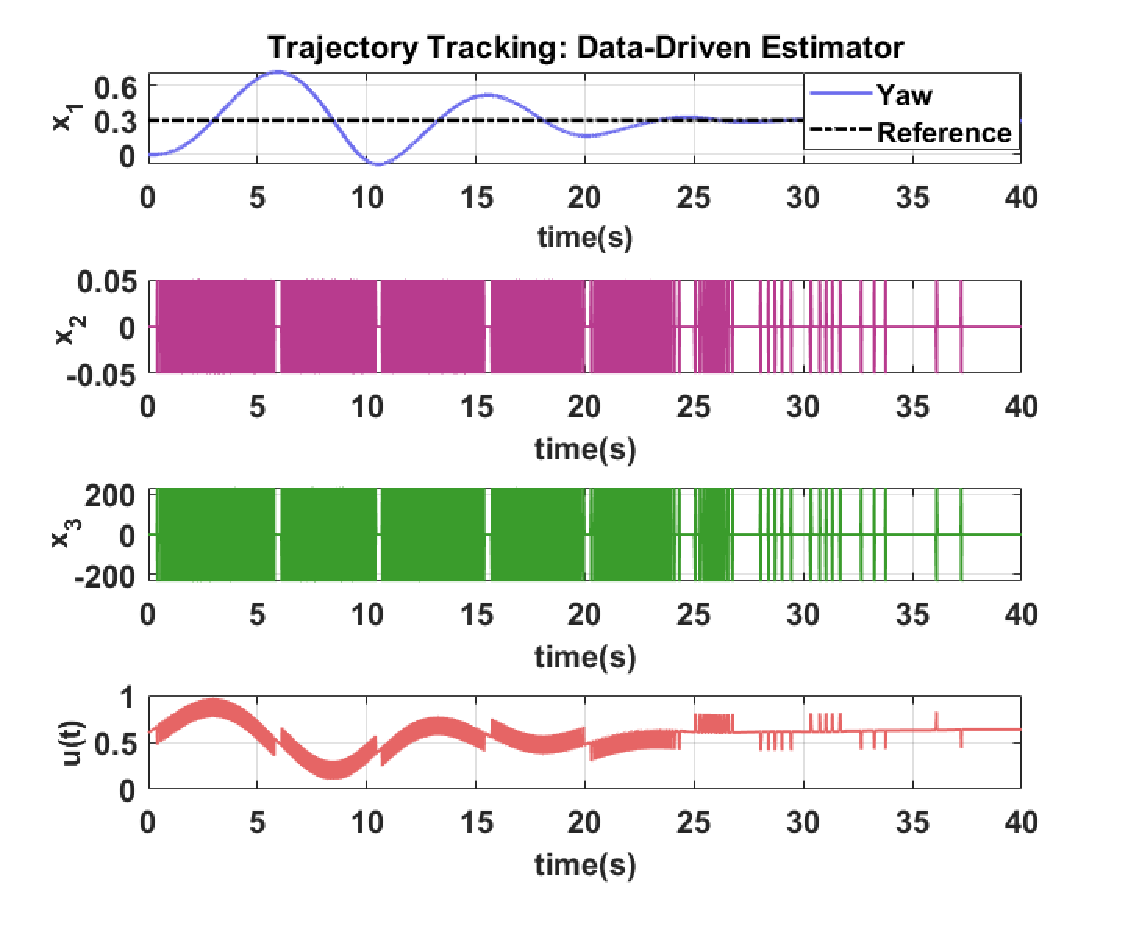}
                                                              
                          \caption{The plot shows the state estimates and the control input when a data-driven estimator proposed in \cite{tabuada} is used with 9 samples to initialize the estimator.}
      \label{ddc}
   \end{figure}

\begin{figure}
      \centering
      \includegraphics[scale=0.4]{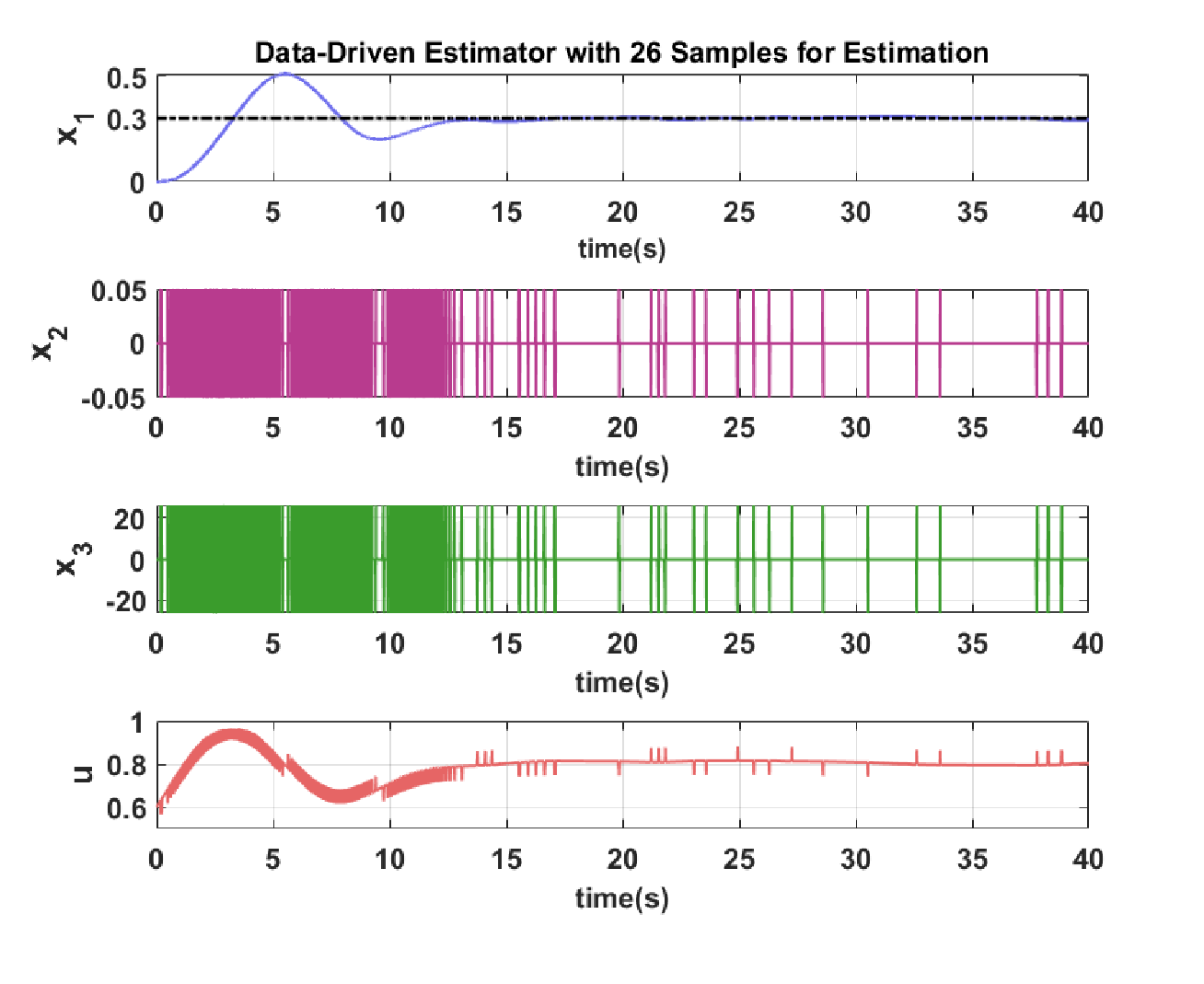}
      \caption{The plot shows the state estimates and the control input when a data-driven estimator, proposed in \cite{tabuada}, is used with 26 samples for initializing the estimator.}
      \label{ddc26}
   \end{figure}

\begin{figure}
      \centering      \includegraphics[scale=0.5]{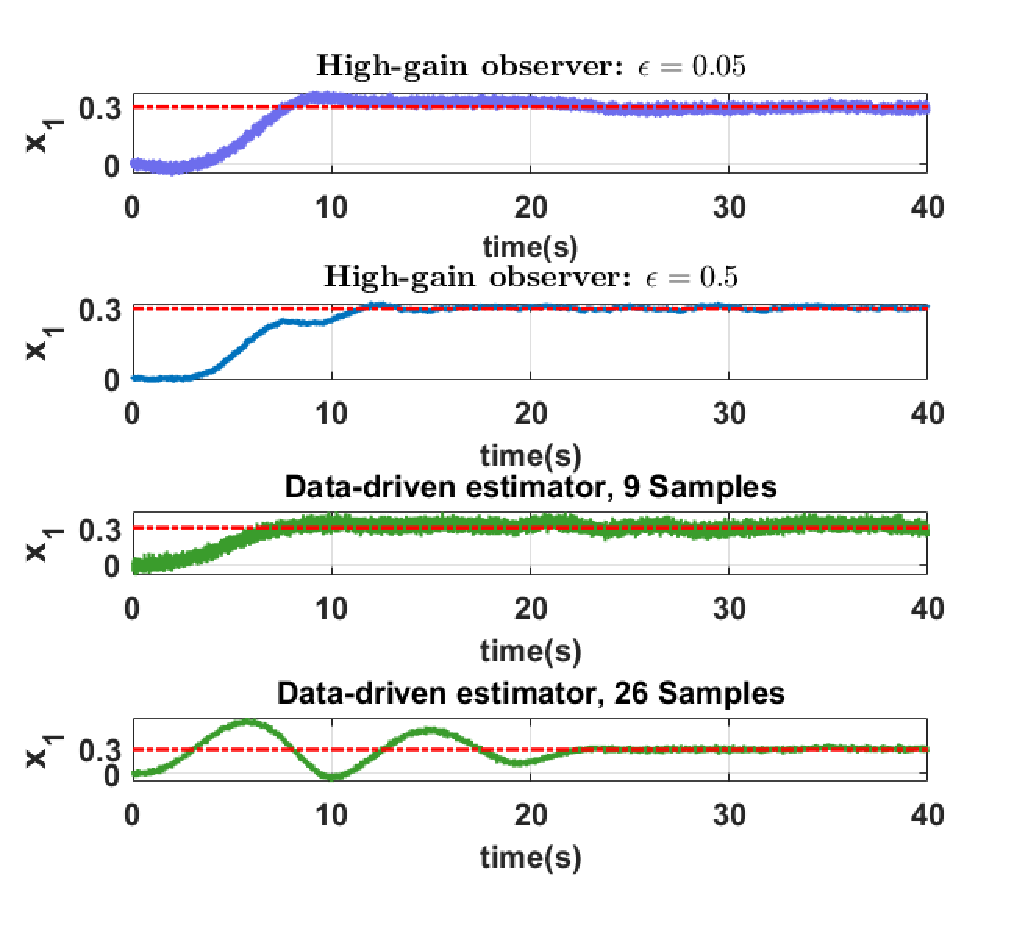}
      \caption{Sensor noise is reduced by an increase in $\epsilon$ for the high-gain observer. In the case of the data-driven estimator, this is achieved by increasing the number of samples used for initializing the estimator.}
      \label{noise}
   \end{figure}

We compare the method proposed with the data-driven technique in terms of ease of implementation and tuning, and robustness in the presence of noise. \\
\textbf{Ease of Implementation and Tuning}: Both methods require a considerable amount of tuning since the model is assumed to be completely unknown. In the case of the high-gain observer, the parameter $\epsilon$ is chosen such that the upper bound on $\epsilon$ is dictated by the stability of the system and the lower bound by the amount of noise the observer rejects. In the data-driven estimator, the number of samples required for estimation is dictated by the number of states and the amount of noise the estimator rejects. The peaking in the states is similar to the peaking phenomena observed in the high-gain observer. \\
In order to compare the two methods with respect to ease of implementation and performance, we replaced the high-gain observer with a data-driven estimator. No other parameters were changed and the number of samples used for estimation was set to 9. The data-driven estimator gave an overshoot of over $100\%$ and large peaking. Fig~\ref{ddc} shows the state estimates and the control input in the case of a data-driven estimator. The overshoot and the peaking are high as only 9 samples were being used for state estimation. Increasing the number of samples will increase the numerical computations required in the estimation. This is shown in Fig~\ref{ddc26}, where the overshoot has reduced to $70\%$ when we use 26 samples for estimation. 
   \\
\textbf{Robustness in the presence of noise: }To test the robustness of both the methods in the presence of noise, an additional noise of 1kHz (since there are built-in filters in the TRMS which suppress noise) with $0$ mean and a variance of $0.001$ is added to the yaw angle sensor output using the random number generator block. All other parameters in both the methods remain unchanged. The high-gain observer with an observer time constant $\epsilon=0.05$ is seen to have better noise attenuation than the data-driven estimator. Moreover, the noise in the output of the high-gain observer reduces significantly when the observer time constant $\epsilon$ is raised to $0.5$. This is because, in the case of the high-gain observer, the states can be shown to converge to a ball of radius proportional to $\epsilon^{-\rho}$. Whereas in the case of a data-driven estimator the radius is proportional to $T^{-\rho}$. Hence, one has to change the sampling time, which is a hardware parameter, or the number of samples used for estimation, which will require a larger computational power, to reduce the bound on the states in the presence of noise when using a data-driven estimator. Thus the advantage of using a high-gain observer, where the performance of the system depends on a software parameter($\epsilon$), is highlighted here. The state trajectories and the control input are given in Fig~\ref{noise}. 

Note that in the model-free control design  required neither the use of the system model, nor apriori collection of open-loop data. 
\subsection{Inferences}
The following inferences are derived from the experimental results.
\begin{itemize}
    \item The high-gain observer with the dynamic controller performed better in terms of system transients as compared to the data-driven method. Using the same controller parameters ($\gamma,\;k_1,\;k_2$), the proposed method resulted in lesser overshoot in the step response of the yaw angle as compared to the method in \cite{tabuada} which uses a data-driven estimator.
    \item Computationally, the high-gain observer can be seen to be much cheaper compared to the data-driven estimator while yielding the same performance. This is because the performance of the data-driven estimator depends on the number of samples used for estimation and this directly translates to the number of computations needed for state estimation(see Equation 6.5 in \cite{tabuada}).
    \item In the presence of sensor noise, both the data-driven estimator and the high-gain observer have low-pass filtering characteristics (see Fig. \ref{noise}). The caveat here is that, in the case of the high-gain observer, the noise can be suppressed by increasing the observer time constant $\epsilon$ (whose upper bound is decided through Theorem \ref{th1} to ensure the stability of the system). Whereas, for the data-driven estimator, the same can be achieved only by increasing the sampling time ($T$) (whose upper bound is decided through Theorem 8.1 and 8.2 of \cite{tabuada} to ensure the stability of the system), or by increasing the number of samples used for estimation. This again increases the computational cost in the latter solution. Note that it is undesirable to increase the number of open-loop data samples used for estimation, especially when stabilizing unstable equilibrium points for hardware plants.
\end{itemize}
\section{CONCLUSIONS}
This article proposes a model-free controller for the stabilization of minimum-phase feedback linearizable nonlinear systems. The method proposed does not require any apriori open-loop data collection for the estimation unlike the recent data-driven techniques, which gives our method the advantage when the hardware plants are open-loop unstable. In our method, we select the high-gain observer for the estimator stage, because of its ease of implementation, lesser computational cost, and superior noise attenuation as compared to the data-driven estimator. This fact was validated using experiments on a twin-rotor system. The proposed method, as compared to the recent data-driven methods, exhibits better performance in terms of overshoot, settling time, and robustness to sensor noise.  The role of sampling time, the number of samples used in data-driven estimation, and the observer time constant in a practical scenario were highlighted. We demonstrated using experiments that increasing the number of samples used in data-driven estimation reduces the overshoot and increases the noise rejection. Lastly, we believe the discussions presented in this article open up further directions for research in this area. In particular, it is worthwhile investigating cases where the zero dynamics exhibit a stable limit cycle and also cases where the input appears in the state equations  governing  the internal dynamics of the system,(meaning the first expression in \eqref{fblinsys} would appear as $\dot w - f_0( w,x, u)$), for example in the pendulum on a cart system. Addressing these issues is a part of our ongoing and future research.

\bibliographystyle{IEEEtran}
\bibliography{ref}

\end{document}